\documentclass[]{aa}
\usepackage{graphicx}
\begin{document}
\title{Possible detection of a magnetic field in T Tauri}

\author{D.A. Smirnov   \inst{1}
   \and S.N. Fabrika   \inst{2}
   \and S.A. Lamzin    \inst{1,3}
   \and G.G. Valyavin    \inst{2}
       }

\offprints{D.\,Smirnov, \\ \email{danila@sai.msu.ru}}

\institute{Sternberg Astronomical Institute, Moscow V-234, 119992, Russia
      \and
	   Special Astrophysical Observatory, Nizhnij Arkhyz, 357147, Russia
      \and
          Isaak Newton Institute of Chile Branch in Moscow, Russia
          }

\date{Received / Accepted }

\abstract{ 
   Medium-resolution $(R\simeq 15000)$ circular spectropolarimetry of
T Tauri is presented. The star was observed twice: on November 11, 1996 and
January 22, 2002. Weak circular polarization has been found in photospheric
absorption lines, indicating a mean surface longitudinal magnetic field
$B_{\|}$ of $160\pm 40$ G and $140\pm 50$ G at the epoch of the first and
second observations respectively. While these values are near the detection
limit of our apparatus, we belive that they are real. In any case one can
conclude from our data that $B_{\|}$ of T Tau does not significantly exceed
200 G, which is much less than surface magnetic field strength of the star
($>2.3$ kG) found by Guenther et~al. (1999) and Johns-Krull et~al. (2000).
We discuss possible reasons of this difference.
         }
   \maketitle

\section{Introduction}

 Classical T Tauri stars (CTTSs) are low mass pre-main sequence stars which
have accretion disks (Bertout, 1989). Magnetic fields are belived to play a
crucial role in the evolution of CTTSs angular momentum and in the
interaction between the central star and the circumstellar matter. Surface
magnetic fields $\sim 1$ kG are expected from theoretical arguments for
these objects -- see e.g. Shu et~al. (2000) and references therein.

  Meanwhile, no field has been directly detected in CTTSs until recently
although several attempts have been made (Babcock 1958; Brown \& Landstreet
1981; Johnstone \& Penston 1986,1987). Recently two groups independently
succeded in detecting magnetic fields on T Tau, LkCa 15 (Guenther et~al.,
1999; Johns-Krull et~al., 2000) and BP Tau (Johns-Krull et~al., 1999a,b).

   Two general methods have been used to measure the magnetic field of T
Tau. The first one is based on the fact that magnetic fields produce an
enhancement of line equivalent widths that vary with magnetic sensitivity
(Lande factor and line splitting pattern). With this technique one can
derive the product of the surface magnetic field strength, $B$, and surface
filling factor $f,$ i.e. $B\cdot f$-value -- see Basri et~al. (1992) for
details. In the case of T Tau, for example, Guenther et~al. (1999) found
$B\cdot f=2.35\pm 0.15$ kG, which means the average value of $B$  itself
is $\ge$ 2.3 kG.

  The other direct method for detecting magnetic fields is to look for net
circular polarization in Zeeman-sensitive lines. Generally, Zeeman $\sigma$
components are elliptically polarized, with the components of opposite
helicity split to either side of the nominal wavelength $\lambda_0.$ If
there is a net longitudinal component of the magnetic field 
on the stellar surface, net polarization results in a shift between
Zeeman-sensitive lines observed in right- and left-circularly polarized
(RCP and LCP) light. The shift is
$$
\Delta \lambda_{rl} \simeq 
9.3\cdot 10^{-7} g\, \lambda_0^2\, B_{\|},\,\, \mbox{ m\AA}
\eqno(1)
$$
where $g$ is the Lande factor of the transition, which we  adopt from the
VALD database (Kupka at al. 1999). Here the mean longitudinal magnetic field
averaged over the star including limb darkening, $B_{\|}$, and $\lambda_0$ are
expressed in kG and \AA\, respectively.

  Only upper limits of $B_{\|}$ were found in the case of T Tau with this
technique: 1000 G (Babcock, 1958) and 810 G (Brown \& Landstreet, 1981) at
the $3\sigma$ level. We present here results of our more sensitive
spectropolarimetric measurements of T Tau's longitudinal magnetic field. The
following set of T Tau's atmospheric parameters were adopted: $T_{eff}\simeq
5250 $ K, (K0\,V spectral type), $\log g =3.73$ (White \& Ghez, 2001) and
solar metallicity.


\section{Observations}

  We observed T Tau with the 6\,m telescope of the Special Astrophysical
Observatory on November 21, 1996 and January 22, 2002. The Main Stellar
Spectrograph (MSS - Panchuk, 2001) equipped with the polarimetric analyzer,
(Najdenov \& Chountonov, 1976; Chountonov, 1997) was used. The analyzer
splits the incoming beam into two parallel beams that are offset from one
another along the spectrograph slit  by 5 arcseconds. One beam contains the RCP component of
the spectrum and the other the LCP component. As a result, two spectra with
opposite circular polarization $\sim 200$ \AA\, in length were projected
onto an $1160 \times 1024$ pixel CCD. The spectral band $6300 - 6500$ \AA\,
was observed in 1996 and $6600 - 6800$ \AA\, in 2002. The width of MSS's
slit was $0.5\arcsec$ providing a 2-pixel spectral resolution of $R=\lambda
/\Delta \lambda \simeq 15000.$

  Spectra were processed as follows. Dark current, night sky emission and
detector bias as well as cosmic ray traces, were removed in a standard way,
using routines from the MIDAS software package. A spectrum of a
thorium-argon lamp was used for wavelength calibration. Then we identifed
spectral lines adopting line information from the VALD database (Kupka
at~al., 1999). The shifts between positions of identical lines in spectra
with opposite circular polarization in the same image, $\Delta
\lambda_{rl}$, were derived by means of a Fourier cross-corellation method.
The method appears to be relatively insensitive to uncertainties in the
continuum level position and gives the ability to evaluate the displacement
of line profiles as a whole (Klochkova et~al., 1996).

  We took into account systematic instrumental effects (SIEs) in different
ways reducing the 1996 and 2002 observations. In the first case, we used
observations of the G9.5 III star $\varepsilon$ Tau, which presumbly has
negligible field. Specifically, we assumed that nonzero values of
$\Delta \lambda_{rl}(\lambda)$ in $\varepsilon$ Tau's spectrum result from
instrumental effects only and used them to correct $\Delta \lambda_
{rl}$-values in the spectrum of T Tau -- see Romanyuk et~al. (1998) for more
details. For this reason, our November 21, 1996 observations were organized
in the following way: initially we observed $\varepsilon$ Tau, then T Tau
and then $\varepsilon$ Tau again. Approximate signal-to-noise ratio (S/N) of
individual spectra, number of observed spectra, $n,$ visual magnitude V, and
total exposure time, $t_e$ (in seconds), for each target are presented in
Table 1.

\begin{table}[h!]
\caption[]{Journal of Observations}
\begin{flushleft}
\begin{tabular}{llllll}
\noalign{\smallskip}
\hline
\noalign{\smallskip}
Date & Target & V & n & t$_e$ & S/N \\
\noalign{\smallskip}
\hline
\noalign{\smallskip}
22.11.1996 & $\varepsilon$ Tau  & 3.5 & 5 & 220  & 300 \\
           & T Tau              & 9.9 & 3 & 3600 & 150 \\
21.01.2002 & 75 Tau             & 5.0 & 3 & 1320 & 300 \\
           & $\varepsilon$ Tau  & 3.5 & 3 & 600  & 300 \\
           & T Tau              & 9.9 & 3 & 7200 & 100 \\
24.01.2002 & HD 30466           & 7.3 & 1 & 1400 & 200 \\
\noalign{\smallskip}
\hline
\end{tabular}
\end{flushleft}
\end{table}

  In the case of the 2002 observations, SIEs were removed as follows. Between
exposures the phase compensator was advanced 1/2 wave in order to reverse
the sense of circular polarization recorded in the two spectra.
For each target, the wavelength of a given line in the RCP and LCP spectra
is recorded in the first exposure $(\lambda^r_1, \,\,\lambda^l_1)$ and in
the second $(\lambda^r_2,\,\,\lambda^l_2).$ Then we calculated
$$
\Delta \lambda_{rl} \equiv \lambda^r-\lambda^l = 
{\left( \lambda^r_1-\lambda^l_1 \right) + 
\left( \lambda^r_2-\lambda^l_2 \right) \over 2},
$$
which is expected to be free from the main instrumental effects -- see
Johns-Krull et~al. (1999b) for details. 

  Information on the January 2002 observations is also presented in Table 1,
where now $n$ stands for the number of pairs of individual observations of
any particular star, between which the phase compensator was rotated.


\section{Results}

\subsection{1996 observations}

   Figure 1 presents values of $\Delta \lambda_{rl}(\lambda)$ derived from all
the $\varepsilon$ Tau spectra observed immediately before and after
observations of T Tau. The $\Delta \lambda_{rl}(\lambda)$ dependence was
approximated with a linear function $\Delta \lambda_{rl}=a\cdot N + b,$
where $N$ is the pixel number, for both observations. We found 
$a=(-2.8\pm 2.6)\cdot 10^{-3}$ m\AA\, per pixel, $b=10.5\pm 1.4$ m\AA\, for
the open squares in Figure 1 and $a=(-5.3\pm 3.4)\cdot 10^{-3}$ m\AA\, per
pixel, $b=14.4\pm 1.9$ m\AA\, for black triangles in Figure 1.

  The difference between the parameters of the fits for both datasets
(squares and tringles) is whithin $2.5\,\sigma,$ i.e. is not significant
from  a statistical viewpoint. So one can use the resulting linear
function, derived from all points, to remove SIEs from $\Delta \lambda_{rl}
$-values in the spectra of T Tau. We found $a=(-4.1\pm 2.2)\cdot 10^{-3}$ 
m\AA\, per pixel, $b=12.4\pm 1.2$ m\AA\, in this case -- solid
line in Fig.1

\begin{figure}[h!]
 \begin{center}
  \resizebox{7cm}{!}{\includegraphics{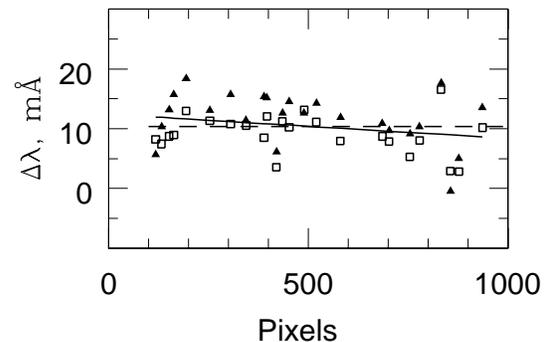}}
  \caption{$\Delta \lambda_{rl}$-values measured in $\varepsilon$ Tau's 
           spectra observed before (filled tiangles) and after (open
           squares) T Tau. See text for details.}
 \end{center}
\end{figure}

  But the $a$-value differs from zero  by less than 2$\sigma,$ 
which means that it is possible that the SIEs do not vary along the CCD. 
With this hypothesis we found the mean shifts are $11.8\pm 0.9,$ and $9.1\pm
0.7$ m\AA\, for filled triangles and open squares respectively. The mean
shift $C$ derived for all points is $10.4$ m\AA\, -- dashed line in Fig.1 --
and standard error is $\sigma_{C}=0.8$ m\AA, or $\sigma_{C}=21$ G, assuming
$g=1$ and $\lambda_0=6400$ \AA\, according to equation (1). The mean shifts
for filled triangles and open squares differ from $C$ less than
$2\sigma_C.$ So it  appears resonable to use  the mean $C$ to remove SIEs,
subtracting it from the measured wavelength shifts between right and left
circularly polarized spectra of T Tau.

  The corrected $\Delta \lambda_{rl}$-values in T Tau's spectra were
used to calculate $B_{\|}$ from equation (1). Resulting values of $B_{\|}$ are
presented in Table 2 for each of three observed spectra -- the last three
columns of the table. In some cases, a given line profile was distored due
to cosmic ray events and thus was not used; these cases are marked with a
dash. In fact, almost all absorption features in T Tau's spectra are
blends of some number of lines, given in the first two columns. So we use an
effective Lande factor, $g_{eff}$: 
$$
g_{eff} = {
\Sigma \, d_i \, g_i
\over \Sigma \, d_i
},
$$
derived from averaging over individual lines which form the blend. Here
$g_i$ is the Lande factor of each line and $d_i$ is its central depth. This
approach appears resonable as long as the Zeeman splitting is less than the
spectral resolution of our instrument. We adopted T Tau's $d_i$-values from
the VALD database specifing $\log\,g$ and $T_{eff}$ of the star as input
parameters. Averaging over all lines gives $B_{\|}=+162$ G with standard
error $\sigma_B = 30$ G  due to measurment uncertainty. Adding the
error due to the uncertainty in $C$, the total uncertainty is:
$$
\sigma = \sqrt{ \sigma_C^2 + \sigma_B^2 } \simeq 37\,\rm{ G.}
$$
Thus, we estimate $B_{\|}=+160\pm 40$ G from our 1996 observations.


\begin{table}[h]
\caption{ Results of 21.11.1996 observations of T Tau }
\begin{flushleft}
\begin{tabular}{cccccc}
\noalign{}
\hline
\noalign{}
Ion & $\lambda$  & g$_{eff}$ & B$_1$ & B$_2$ & B$_3$ \\
\noalign{}
\hline
\noalign{\smallskip}
 Ni {\sc i} & 6314.65 &   1.18 & $-$ & $+$203 & $+$11 \\
 Ni {\sc i} & 6314.66 &  & & & \\ 
 Fe {\sc i} & 6315.31 &  & & & \\
 Fe {\sc i} & 6315.42 &  & & & \\
 Fe {\sc i} & 6315.81 &  & & & \\
\noalign{\smallskip}
 Fe {\sc i} & 6318.02 &  0.72 & $-$ & $+$638 & $-$159 \\
 Ca {\sc i} & 6318.11 & & & & \\
\noalign{\smallskip}
 Ni {\sc i} & 6322.16 &  1.39 & $-$ & $+$251 & $+$88 \\
 Fe {\sc i} & 6322.68 &  & & & \\
\noalign{\smallskip}
 Ni {\sc i} & 6327.59 &  1.05 & $-$14 & $+$157 & $-$ \\
\noalign{\smallskip}
 Cr {\sc i} & 6330.09 &  1.64 &  $+$262 & $-$36 & $+$177 \\
 Fe {\sc i} & 6330.85 &  & & & \\
\noalign{\smallskip}
 Fe {\sc i} & 6338.88 &   1.23 & $+$500 & $+$200 & $-$123 \\
 Ni {\sc i} & 6339.11 &  & & & \\
\noalign{\smallskip}
 Ca {\sc i} & 6343.31 &   1.09 &  $+$18 & $-$ & $+$321 \\
 Fe {\sc i} & 6344.15 & & & & \\
\noalign{\smallskip}
 Fe {\sc i} & 6355.03 &  0.98 & $+$470 & $+$110 & $-$ \\
\noalign{\smallskip}
 Fe {\sc i} & 6358.63 &  1.49 & $+$198 & $+$35 & $+$242 \\
 Fe {\sc i} & 6358.70 &  &&& \\
\noalign{\smallskip}
 Fe {\sc i} & 6393.60 &  1.15 &  $+$288 & $-$46 & $+$128 \\
 Fe {\sc i} & 6392.54 & &&& \\
\noalign{\smallskip}
 Fe {\sc i} & 6400.00 & 1.35   & $+$28 & $+$365 & $-$26 \\
 Fe {\sc i} & 6400.32 & &&& \\
\noalign{\smallskip}
 Fe {\sc i} & 6411.11 &  1.49  & $+$162 &$-$ & $+$248 \\
 Fe {\sc i} & 6411.65 & &&& \\
 Fe {\sc i} & 6412.20 & &&& \\
\noalign{\smallskip}
 Fe {\sc i} & 6430.85 &  1.24 &  $-$153 & $+$399 & $-$215 \\
\noalign{\smallskip}
 Ca {\sc i} & 6439.08 &  1.12 & $+$210 & $+$254 & $+$313 \\
\noalign{\smallskip}
 Ca {\sc i} & 6449.81  & 1.10  & $+$48  & $+$8 & $+$143 \\
 Co {\sc i} & 6450.08 & &&& \\
 Co {\sc i} & 6450.25 & &&& \\
\noalign{\smallskip}
 Ca {\sc i} & 6462.57 &  1.17  & $+$588 & $+$124 & $-158$ \\
 Fe {\sc i} & 6462.71 & &&&\\ 
 Fe {\sc i} & 6462.73 & &&&\\
\noalign{\smallskip}
 Ca {\sc i} & 6471.66 &  1.20 & $+$319 & $-$53 & $+342$ \\
\noalign{\smallskip}
 Ca {\sc i} & 6475.24  & 1.63 & $+$105 & $+$232 & $+$430 \\
 Fe {\sc i} & 6475.62 & &&& \\
\noalign{\smallskip}

\noalign{\smallskip}
\hline
\end{tabular}
\end{flushleft}
\end{table}



\subsection{2002 observations}

   The $\Delta \lambda_{rl}(\lambda)$-values for all sets of $\varepsilon$
Tau's spectra are shown in Figure 2. It can be seen that there is
no statistically significant trend in these data around  an average value
$\Delta \lambda_{rl} = -0.02 \pm 0.32$ m\AA. Assuming $g_{eff}=1$ and
$\lambda_0=6700$ \AA\, (the average wavelength of the spectral
band), this value corresponds to $B_{\|}=-0.5\pm 8$ G. Small values ($\Delta
\lambda_{rl} = -0.93 \pm 0.43$ m\AA\, or $B_{\|}= -20 \pm 10$ G) were also
found for an additional giant star 75 Tau (K2\,III). Thus, we conclude that
the adopted method of SIEs removal works well and confirmed {\it
a'posteriori} our choice of $\varepsilon$ Tau as a zero-field standard
during the 1996 observations.

\begin{figure}[h!]
 \begin{center}
  \resizebox{7cm}{!}{\includegraphics{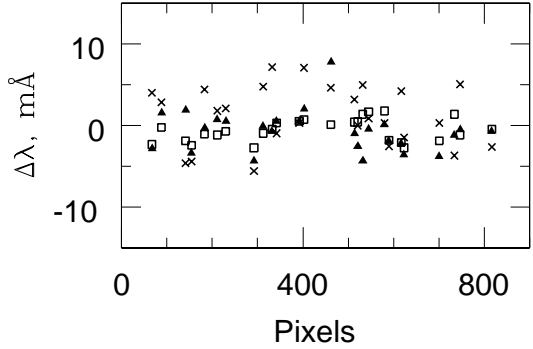}}
  \caption{$\Delta \lambda_{rl}$-values derived from three pairs of 
           $\varepsilon$ Tau's spectra observed in 22.01.2002. 
           See text for details.}
  \end{center}
\end{figure}

  To illustrate the reliability of our technique we included the magnetic
star HD30466 in the target list of our 2002 observations. Babcock (1958)
found $B_{\|}= +2320\pm 340$ G and $B_{\|}= +1890\pm 130$ G for HD30466 from
two independent observations. Profiles of the Cr {\sc i} 6661.08 line in our
HD30466's RCP and LCP spectra are shown in Fig.3: they are clearly shifted
relative each other $(\Delta \lambda_{rl} \simeq 126$ m\AA), indicating
$B_{\|} \simeq +2.0$ kG, adopting $g=1.48$ (Kupka et~al. 1999)

\begin{figure}[h!]
 \begin{center}
  \resizebox{7cm}{!}{\includegraphics{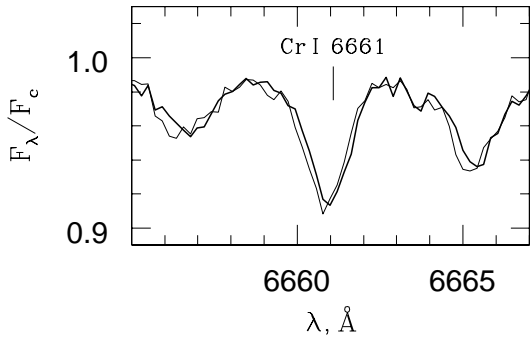}}
  \caption{A part of RCP (thin line) and LCP (solid line) spectra of
           HD30466}
  \end{center}
\end{figure}

  As can be seen from Table 3, relative shifts of lines between spectra with
opposite polarizations in T Tau's spectrum are much less, and we find
$B_{\|}=+140\pm 50$ G after averaging over all the data. The
uncertainty is calculated by taking the standard error of the individual
points.


\begin{table}[h]
\caption{ Results of 22.02.2002 observations of T Tau }
\begin{flushleft}
\begin{tabular}{cccccc}
\noalign{\smallskip}
\hline
\noalign{\smallskip}
Ion & $\lambda$  & g$_{eff}$ & B$_1$ & B$_2$ & B$_3$ \\
\noalign{\smallskip}
\hline
\noalign{\smallskip}
 Fe {\sc i} & 6609.1 & 1.59  & $-$161	& $-$33	 & $-$76  \\
 Fe {\sc i} & 6609.7 &       &          &        &        \\
\noalign{\smallskip}
 V {\sc i}  & 6624.8 & 0.94  & $+$768	& $-$21	 & $+$918 \\
 Fe {\sc i} & 6625.0 &	     &	        &        &        \\
\noalign{\smallskip}
 Fe {\sc i} & 6633.4 & 1.48  & $+$576	& $+$218 & $-$299 \\
 Fe {\sc i} & 6633.8 &	     &	        &        &        \\
 Fe {\sc i} & 6634.1 &       &     	&	 &        \\
\noalign{\smallskip}
 Ni {\sc i} & 6643.6 & 1.31  &  $-$	& $-$8	 & $-$182 \\
\noalign{\smallskip}
 Fe {\sc i} & 6663.2 & 1.53  & $+$160	& $+$58	 & $+$460 \\
 Fe {\sc i} & 6663.4 &       &	        &        &        \\
\noalign{\smallskip}
 Fe {\sc i} & 6677.9  & 1.35 & $+$143	& $-$207 & $-$67  \\
 Fe {\sc i} & 6678.0  &	     &	        &        &        \\
\noalign{\smallskip}
 Al {\sc i} & 6696.0 &  0.97 & $-$101	& $-$182 & $+$545 \\
 Fe {\sc i} & 6696.3 &       &          &        &        \\
\noalign{\smallskip}
 Al {\sc i} & 6698.7 & 1.25  & $+$198	& $+$633 & $-$214 \\
 Fe {\sc i} & 6699.1 &       &	        &        &        \\
\noalign{\smallskip}
 Li {\sc i} & 6707.8 & 1.25  & $+$79	& $+$91	 & $+$87  \\
\noalign{\smallskip}
 Fe {\sc i} & 6710.3 & 1.69  & $-$82	& $-$162 & $-$150 \\
\noalign{\smallskip}
 Ca {\sc i} & 6717.7 & 1.01  &  $+$7    & $+$52  & $+$232 \\
\noalign{\smallskip}
 Ti {\sc i} & 6743.1 & 1.01  & $+$735	& $+$633 & $+$635 \\
\noalign{\smallskip}
 Fe {\sc i} & 6750.2 & 1.50  & $-$201	& $-$77	 & $+$736 \\
\noalign{\smallskip}
 Fe {\sc i} & 6752.7 & 2.01  & $-$282	& $-$161 & $+$61  \\
\noalign{\smallskip}
 Ni {\sc i} & 6767.8 & 1.43  & $+$192	& $-$111 & $+$646 \\
\noalign{\smallskip}
\hline
\end{tabular}
\end{flushleft}
\end{table}


\section{Discussion}

   From observations in 1996 and 2002, using different methods of SIE
correction, we find similar $(+160\pm 40$ G and $+140\pm 50$ G) values of
the mean surface longitudinal magnetic field in T Tau.  As far as such
small field values are close to the detection limit, it is very important to
show that the measured wavelength shift is magnetic in origin. To test this
we have measured relative shifts of the S\,{\sc ii} 6716.44 \AA\, and
6730.82 \AA\, forbidden emission lines in our 2002 spectra, which presumbly
originate far from the star in a region with (nearly) zero magnetic field.
We found $\Delta \lambda_{rl} = +0.8 \pm 1.4$ m\AA\, for these lines which
would give $20\pm 35$ G for $g_{eff}=1.$ This result can be considered as
evidence that the measured wavelength shift of the stellar absorption lines
is magnetic in origin.

  To further test if it is indeed possible to detect such low $B_{\|}$ value
from our spectra we performed the following test. The RCP and LCP spectra of
1D stellar atmosphere were calculated with the SYNTHMAG program (Piskunov,
1998), assuming that $\log g$ and $T_{eff}$ are equal to that of T Tau and
$B=B_{\|}=200$ G, i.e. with magnetic field lines perpendecular to the
 stellar surface. Profiles of lines within 6300-6500 \AA\, spectral band were
artifically broadened to take into account stellar rotation and spectral
resolution  and binned  to the appropriate dispersion. Each of these
artifical specra were used to produce three new ones by means of adding
noise at the appropriate level. Finally we derived three pairs of RCP and
LCP spectra with S/N-ratio near 100, as in the case of our 1996 and/or 2002
observations. By means of cross-corellation of RCP and LCP profiles we then
found $B_{\|}=164\pm 32$ G in agreement with the initial value of 200 G.
Therefore, we belive that we indeed have recorded T Tau's surface magnetic
field $B_{\|} \sim 150$ G, in spite of the relatively large scatter of the
individual mesurements -- see Table 2\& 3. Regardless, our observations show
T Tau's large scale $B_{\|}$-value did not significantly exceed 200 G.

  This value is much less than the average value of T Tau's surface magnetic
field (B$> 2$ kG) found by Guenther et~al. (1999) and Johns-Krull et~al.
(2000). This will result if there are regions with opposite polarity of
magnetic field lines at the stellar surface. This could indicate the
magnetic field of T Tau is non-dipolar. Note however, that even in the case
of a magnetic dipole, the observed $B_{\|}=0$, if the dipole's axis is
perpendecular to the line of sight. To demonstrate this more clearly we
considered a magnetic dipole with equatorial field strength B$=1.6$ kG and
calculated using the SYNTHMAG program the relative shifts of RCP and LCP
profiles of the Fe\, {\sc i} 6430.8 line for different inclination angles
$\beta$ and then derived the "observed" longitudinal field. We found
$B_{\|}\simeq 900,$ 400 and 200 G for $\beta=0\degr,$ $60\degr$ and
$75\degr$ respectively. The mean unsigned surface field  values for these
inclinations are 2.4, 2.2 and 2.1 kG respectively, i.e. close to values
found by Guenther et~al. (1999) and Johns-Krull et~al. (2000).

  T Tau is believed to have a low inclination to its rotation axis based on
$v\sin\,i$ (Hartmann \& Stauffer, 1989) and rotation period measurements
(Herbst et~al., 1985). Therefore, one can expect that the magnetic axis is
also close to the line of sight if it is aligned with or close to the
rotation axis. Thus the low value of the average longitudinal magnetic field
of T Tau we found can be interpreted in two possible ways: 1) as an
indication of the non-dipole character of the stellar magnetic field, or 2)
as evidence of a large $(\ga 30\degr)$ angle between the rotation and
magnetic axes.


\section{Conclusion}

  Weak circular polarization has been found in photospheric absorption lines
of T Tau, indicating a mean longitudinal magnetic field $B_{\|} \sim
150$ G. While this value is near the detection limit of our apparatus,
we belive that it is real. In any case one can conclude from our data that
$B_{\|}$ on T Tau does not significantly exceed 200 G, i.e. much less
than surface magnetic field strength of the star ($>2.3$ kG) found by Guenther
et~al. (1999) and Johns-Krull et~al. (2000). This could be the result of
the non-dipole character of the magnetic field on T Tau or evidence of a large
angle between the rotation and magnetic axes. More precise observations
are necessary to distinguish between these possibilities.


\begin{acknowledgements}
  
  We thank our colleagues G.A.\,Chountonov, N.E.\,Piskunov,
T.A.\,Ryabchikova, D.N.\,Monin and A.A.\,Panferov for assistance with
observations and data reduction as well as helpful discussions. It is a
pleasure to thank the referee C.M.\,Johns-Krull for a careful reading
of the manuscript and very useful and benevolent comments. Finally we thank
the Russian Fund of Fundamental Research for the grants 02-02-16070 (D.S.
and S.L) and 01-02-16808 (S.F.).

\end{acknowledgements}


\end{document}